Su dongcai

# Data recovery from grossly corrupted observation via l1 minimization[1]

Dongcai Su at 2015-9-22

## Abstract

**This paper studies the problem of recovering a signal vector and the corrupted noise vector from a collection of corrupted linear measurements through the solution of a l1 minimization (2.1.2), where the sensing matrix is a partial Fourier matrix whose rows are selected randomly and uniformly from rows of a full Fourier matrix. After choosing the parameter in (2.1.2) appropriately, we show that the recovery can be successful even when a constant fraction of the measurement are arbitrarily corrupted, provided that the signal vector is sparse enough. The upper-bound on the sparsity of the signal vector required in this paper is asymptotically optimal when the sensing matrix is partial bounded orthonormal system (BOS) [1] and is better than those achieved by recent literatures [2, 3] by a $\ln(n)$ factor. Furthermore, the assumptions we impose on the signal vector and the corrupted noise vector are loosest comparing to the existing literatures [2-4] which lenders our recovery guarantees are more applicable. Extensive numerical experiments based on synthesis as well as real world data are presented to verify the conclusion of the proposed theorem and to demonstrate the potential of the l1 minimization framework.**

# 1. Introduction

Compressed sensing have been well studied [5-7] recently as a novel data sampling framework, which have attracted great attentions in scientific areas [8] as well as industrial applications [9-13]. In a CS sampling framework, a sparse n-dimensional signal $x^{(0)}$ is recovered from a collection of linear projections $b = Ax^{(0)}$, where A is a $m \times n$ sensing matrix , typically, the number of measurement m is significantly smaller than the dimension of $x^{(0)}$ n, to recover $x^{(0)}$, a standard $\ell_1$ minimization is often adopted in the CS literature:

$$\min_x \|x\|_1, \text{s.t.} Ax = b \qquad (1.1)$$

Typical theoretical results in the CS literatures show that if A obeys some properties, e.g., the restricted isometry property (RIP) [14, 15], then the solution of (1.1) can recover $x^{(0)}$ with high probability provided that $x^{(0)}$ is sparse enough [16].

It is now well-known that RIP is satisfied with high probability for a broad family of sensing

---

[1] Part of this paper had been submitted to *The* ***2016*** *IEEE* ***International Symposium on Information Theory (isit 2016)***

matrices, e.g., when A is the sub-gaussian matrix [1, 8, 17], then A satisfies the RIP with high probability if $m \geq O\left(k\ln\left(\frac{en}{k}\right)\right)$. Similarly, when A is the partial bounded orthogonal system (BOS)[1], then A satisfies the RIP as long as $m \geq O(k\ln^4(n))$ [17], where $k \geq \max\left\{1, \left\|x^{(0)}\right\|_0\right\}$.

In many applications, the measurement vector b may also contain certain amount of dense, bounded noise, which is denoted by a m-dimensional vector $v$, and b is expressed as: $b = Ax + v$, with $\|v\|_2 \leq \eta$. To recover $x^{(0)}$ in this case, a convex optimization is usually adopted [16]:

$$\min_x \|x\|_1, \text{s.t.} \|Ax - b\|_2 \leq \eta \qquad (1.2)$$

It can be proved that if A satisfies the RIP, then stable recovery[2] of $x^{(0)}$ by (1.2) is still possible based on the same number of measurements as in (1.1) [14, 18-21]. Alternatively, similar exact or stable recovery results corresponding to (1.1~1.2)can also be obtained through other algorithms such as orthogonal matching pursuit (OMP) [22, 23], and the analysis is based on the mutual coherence property of the sensing matrix, which is closely related to the RIP chap 5~6 of [1].

Generally speaking, it require asymptotically more number of measurement to guarantee the RIP of A or successful recovery of (1.2) when the sensing matrix A is partial BOS than when A is Gaussian random matrix. Chap. 12 of [1].

However, as pointed out in chap.12 [1], manipulating a completely random sensing matrix might add extra storage or computation burden, for example, storing a partial Fourier basis as the sensing matrix requires far more less memories than storing a random i.i.d Gaussian matrix with the same size, besides, algorithms (e.g. alternating direction multipliers method (ADMM) [24] and Bregman method [25]) solving the $\ell_1$ minimization (1.1) with A as the partial Fourier basis matrix can be much faster than in general case.

Furthermore, as commented by [26, 27], completely random sensing matrix might not be relevant in real applications, for instance, the data acquisition systems in many practical applications are inherently frequent based, in other words, the sensing matrices corresponding to these data acquisition systems are essential partial Fourier basis matrices, we should just mentioned a few of important applications lie in this case as stated below:

- Magnetic Resonance Imaging (MRI). Chap. 1 of [1] [9-12]. Let $\Omega \subset R^3$ be the interested region for imaging, the magnetization at position $z \in \Omega$ is denoted by $X(z) = |X(z)|\exp(-2\pi i\langle k(t), z\rangle)$, where $|X(z)| > 0$ denotes the magnitude, $k(t) = [k_1(t), k_2(t), k_3(t)]^T \in R^3$ is a time dependent vector and $\langle \cdot,\cdot \rangle$ denotes the inner product. Then at time t, the receiver coil integrates $X(z)$ over $\Omega$ and measures the value $f(t) = \iint_\Omega |X(z)|\exp(-2\pi i\langle k(t), z\rangle)dz$. To write it in a discretization way, firstly, we represents $\Omega$ by $N_1 \times N_2 \times N_3 = N$ voxels, let $x \in R^N$ denotes the vector to be reconstructed where $x(j), 1 \le j \le N$ denotes the value of $|X(z)|$ of the corresponding voxel. Secondly, we also we also sample $k(t) = [k_1(t), k_2(t), k_3(t)]^T$ randomly and uniformly within a region: $0 \le k_j(t) < N_j, 1 \le j \le 3$ and $t \in [1, T]$ be a integer. In this manner, one has, $b = Ax$. Where $b \in C^T$ denotes the measurement vector with $b(t) = f(t), t \in [1, T]$, and $A \in C^{T \times N}$ is a submatrix whose rows are selected randomly and uniformly from a $N \times N$ Fourier basis.

[2] Here stable recovery means $\left\|\hat{x} - x^{(0)}\right\|_2 \leq C\eta$, where $\hat{x}$ denotes the solution of (1.2), C is some positive constant.

- Radar. In this application, at time t, a source antenna sends out a properly designed electromagnetic wave, then a receive antenna measures the electromagnetic signal reflected by objects in the sky. Mathematically, let z denotes object in the sky which is parameterized by d—the distance between the object and the receive antenna, then at time t, the measurement of the receive antenna reads $b(t) = \sum_d |z(d)| \exp(-2\pi i \phi(d, t))$, where $|z(d)|$ denotes the magnitude of z(d), $|z(d_0)| = 0$ if and only if there is no object parameterized by $d = d_0$. $\phi(d, t) = cdf(t)$ denotes the phase translation which depends on d and t, where c is a positive constant depends on factors like wave speed, and f(t) denotes the frequency of the electromagnetic wave sent by the source antenna at time t, after proper rescaling operations, let $d \in [0, N-1]$ be integer, let $x \in R^N$ be a vector whose element denotes the value $|z(d)|$ of the corresponding objects. Then we have: $b(t) = \sum_d |z(d)| \exp\left(-\frac{2\pi i d f(t)}{N}\right)$, where $t \in [1, T]$ are integers, if we choose f(t) randomly and uniformly within a given region, then one can express the measurement vector $b \in C^T$ compactly as: $b = Ax$, where A is a $T \times N$ matrix whose rows are selected uniformly at random from a $N \times N$ Fourier basis.
- Planar X-ray. [28] Let $\Omega \subset R^3$ denotes the interested region, let $f(z) \in R, z \in \Omega$ represents the X-ray reaction coefficient of the position z, our goal is to reconstruct $f(z)$. A plane is defined by $P(d, \gamma(t)) = \{z | z^T \gamma(t) = d\}$, where $\gamma(t) \in R^3$ is a normal vector dependent on t, a radon transform defined on the plane reads: $R_f(d, \gamma(t)) = \iint_{\Omega \cap P(d, \gamma(t))} f(z) dz$. Suppose at time t, we've measured a collection of radon transform defined by $\gamma(t)$: $R(t) = \{R_f(d, \gamma(t)) | L(\gamma(t)) \le d \le U(\gamma(t))\}$, where $L(\gamma(t))$, $U(\gamma(t))$ are the minimum, maximum value of d such that $P(d, \gamma(t)) \cap \Omega \ne \emptyset$. The 1-D Fourier transform of $R_f(d, \gamma(t))$ along the d direction is denoted as $\breve{R}_f(s, \gamma(t))$, $s \in R$. It can be proved that when $f(z)$ satisfies certain conditions, $\breve{R}_f(s, \gamma(t)) = \breve{f}(s\gamma(t))$ holds, where $\breve{f}(s\gamma)$, $\gamma \in R^3$ denotes the 3D Fourier transform of $f(z), z \in R^3$. In discretization case, if one choose $\gamma(t)$ uniformly at random, then one has: $b = Ax$, where b is the measurement vector whose elements are obtained by Fourier transform of the radon transform as described above, A is a wide matrix whose rows are selected randomly and uniformly from a Fourier basis, and x is the vector whose element represent f(z) of the corresponding position z.

These applications motivate us to study the signal recovery when the sensing matrix A is partial Fourier basis.

When A is partial Fourier basis and the measurement vector b contains some noise whose norm is bounded above by $\eta$, there exist elegant recovery guarantee for (1.2): the analysis based on RIP [17] show that, if $m \ge O(k \ln^4(n))$, where k is a positive integral represents the upper-bound of $\|x^{(0)}\|_0$ then (1.2) can faithfully recover $x^{(0)}$ with high probability for all $x^{(0)}$ whose cardinality is smaller than k, this result is often call uniformly (or universally) recovery guarantee. The analysis based on the dual certification which is satisfied by a constructive scheme called the golfing-scheme [1, 29] shows that, if $m \geq O(k\ln(n))$, then stable recovery of $x^{(0)}$ can be succeed with high probability for any fixed $x^{(0)}$, and this result is called the non-uniformly recovery guarantee chap.12 [1].

Although these recovery guarantee results are promising, when the energy of the measurement noise grows larger, the recovery error of (1.2) may also becomes unexpectedly large, which means that a few grossly corrupted measurements can severely degrade the performance of (1.2). Unfortunately, corrupted noise and irrelevant measurements are prevalent in modern applications such as image processing, sensor networks where certain amount of measurements are grossly corrupted due to the factors like hardware flaws and environmental hazards [30-32].

And this motivated another line of works, which is called compressed sensing with corruption [2, 33]. Where the measurement vector is represented as $b = Ax^{(0)} + f^{(0)}$, here, $f^{(0)}$ is a m-dimensional vector representing the corrupted noise, which is often assumed to be sparse and whose non-zero elements can take arbitrary values. Similar to (1.1), the signal $x^{(0)}$ and corrupted noise $f^{(0)}$ are recovered through the below $\ell_1$ minimization:

$$\min_{x,f}\|x\|_1 + \theta\|f\|_1, \text{s.t.}\, Ax + f = b \qquad (1.3)$$

*Existing works on compressed sensing with corruption.* [34, 35] apply (1.3) when $\theta = 1$ and A is Full Fourier basis on the problem of separating the sinus and spikes, the deterministic guarantee for (1.3) based on the coherence of matrices require that the sparsity of both $x^{(0)}$ and $f^{(0)}$ are bounded from above by $O(\sqrt{n})$ as indicated in [36, 37]. Later, Candes [4] shows that when A is full Fourier basis, $\theta = 1$, then exact recovery of $x^{(0)}$ and $f^{(0)}$ is possible when the supports of $x^{(0)}$ and $f^{(0)}$ is uniformly random and both $\|x^{(0)}\|_0$ and $\|f^{(0)}\|_0$ is smaller than $O(n/\sqrt{\ln(n)})$.

Motivated by the problem of face recognition, Wright et.al [30] shows that recovery is possible even when $\|f^{(0)}\|_0$ grows arbitrarily close to m, provided that $\|x^{(0)}\|_0$ is sub-linear smaller than m, however, in their "cross and brouquet model", A is a Gaussian i.i.d designed matrix, which is different from our setting.

More recently, Nguyen [3] and Li [2] shows that when A is a partial BOS (where partial Fourier matrix can be treated as a special case), after choosing $\theta$ appropriately as depending on m and n, then recovery of $x^{(0)}$ and $f^{(0)}$ is possible even when a constant fraction of b is corrupted by $f^{(0)}$, provided that $m \geq O(\|x^{(0)}\|_0 \ln^2(n))$.

Recently, [38] studies the probabilistic recovery guarantee of a more general $\ell_1$ minimization:

$\min_{x,f}\|x\|_1 + \theta\|f\|_1, \text{s.t.} Ax + Bf = b$ (1.3)

Where $b = Ax^{(0)} + Bf^{(0)}$, A and B are general matrices, based on the coherence of matrices A and B, the authors in [38] show that recovery of $x^{(0)}$ and $f^{(0)}$ is possible even when the sparsity of $x^{(0)}$ and $f^{(0)}$ scale linearly to the number of measurement $m$, provided that the signs and supports of $x^{(0)}$ and $f^{(0)}$ satisfy some random assumption.

*Our contributions.* Despite the promising theoretical results, in the proofs of the above mentioned literatures [2-4], all impose some restrictions on $x^{(0)}$ and $f^{(0)}$, e.g., random assumptions on the supports or signs of $x^{(0)}$ and $f^{(0)}$, which are unlikely satisfied in real applications. This motivates the research of this paper, for convenience, we reformulate (1.2) as (1.4) below:

$\min_{x,f}\|x\|_1 + \|f\|_1 \text{s.t.} \lambda\tilde{A}x + f = b$ (1.4)

Where $\tilde{A}$ is a column normalized partial Fourier matrix whose rows are chosen randomly and uniformly from a $n \times n$ Fourier basis, it is easy to observe that solutions of (1.2) and (1.4) can be the same if we appropriately tune $\lambda$ in (1.4) depending on $\theta$ in (1.2).

We show that when $\lambda = 1$, (1.4) can recover $x^{(0)}$ and $f^{(0)}$ when a constant fraction of the measurement vector b are arbitrarily corrupted by $f^{(0)}$, provided that $m \geq O(k\ln(k))$, here $k = \max\left\{\|x^{(0)}\|_0, 1\right\}$, this lower bound for m is better than the asymptotically optimal lower bound on $m$ when the sensing matrix is partial BOS chap. 12 [1]. When $\lambda = 1/\sqrt{\ln(2n/\varepsilon)}$, we show that successful recovery of (1.4) is also possible even when the number of corrupted observation $\|f^{(0)}\|_0$ becomes arbitrarily close the total number of measurement m, provided that $m \geq O(k\ln(n))$, this lower bound of m is asymptotically better than those achieved by recent literatures [2, 3] by a $\ln(n)$ factor.

It's worthy to mention that we impose no assumption of the signs of $x^{(0)}$ or $f^{(0)}$, which lenders our theoretical result to be more applicable in practice. Extensive experiments based on synthesis data as well as real data faithfully validate our analysis result.

*Organization of the paper.* The organization of the remaining of the paper is as follows, section 2 provides our main results—theorem 2.1, section 3 provides the proof roadmap of theorem 2.1, with the supporting lemmas detailed in appendix A.1~A.3 which eventually leads to the final proof of theorem 2.1 as given in A.4. In section 4, we provided the experiments verify the conclusion of theorem 2.1 base on both synthesis data and the real image data obtained from the BSDS 500 database. Finally, section 5 summaries the findings of this paper with conclusions and future works.

# 2. Main result

## 2.1 notations and problem statement

Before the problem statement, we briefly introduce some notations that will be used in this section and throughout the rest of the paper.

*Notations.* Let $[\mathrm{n}] = \{1, \dots, \mathrm{n}\}$ denotes an indices set if n represents a positive integer, and $\mathrm{i} + [\mathrm{n}] = \{\mathrm{i}+1, \dots, \mathrm{i}+\mathrm{n}\}$ denotes an indices set [n] be right shifted by i, if i denotes an integer. Let $\mathrm{s_x}$, $\mathrm{s_f}$ denotes some indices subset of [n], [m] which contain the support of $\mathrm{x}^{(0)}$, $\mathrm{f}^{(0)}$, respectively. $|\mathrm{s_x}|$ denotes the cardinality of $\mathrm{s_x}$, here $|\cdot|$ denotes the cardinality of a set if $\cdot$ represents a set, $|\mathrm{s_f}|$ is interpreted similarly. and let $\sigma_\mathrm{x}, \sigma_\mathrm{f}$ denote the sign[3] vector of $\mathrm{x}^{(0)}(\mathrm{s_x}), \mathrm{f}^{(0)}(\mathrm{s_f})$, respectively. $\mathrm{Z}^+$ denotes the set of all positive integers.

Suppose we have a m-dimensional measurement vector $\mathrm{b}$ which can be written as:

$$\mathrm{b} = \widetilde{\mathrm{A}}\mathrm{x}^{(0)} + \mathrm{f}^{(0)} \qquad (2.1.1)$$

Where $\mathrm{x}^{(0)}$, $\mathrm{f}^{(0)}$ denote the n-dimensional ground-true signal and m-dimensional corrupted noise, respectively. $\widetilde{\mathrm{A}}$ is a $\mathrm{m} \times \mathrm{n}$ column normalized partial Fourier basis: $\widetilde{\mathrm{A}} = \sqrt{\frac{\mathrm{n}}{\mathrm{m}}}\mathrm{F}(\wedge, [\mathrm{n}])$, we use the below definition 2.1 to define $\wedge$ -- the row indices set of the sensing matrix $\widetilde{\mathrm{A}}$.

***Definition 2.1*** *(random subset) An indices set $S = \{s_1, \dots, s_m\}$ with cardinality m is called a random subset of [n], if $S \subseteq [n]$ and S is selected randomly and uniformly from all subset of [n] with cardinality m.*

In this paper, our goal is to recover $\mathrm{x}^{(0)}$, $\mathrm{f}^{(0)}$ through the below $\ell_1$ minimization:

$$\mathrm{minimize}_{\mathrm{x,f}}\ \|\mathrm{x}\|_1 + \|\mathrm{f}\|_1\ \mathrm{s.t.}\ \lambda\widetilde{\mathrm{A}}\mathrm{x} + \mathrm{f} = \mathrm{b} \qquad (2.1.2)$$

E.g., let $\hat{\mathrm{x}}, \hat{\mathrm{f}}$ denote the solution of (2.1.2), then $\lambda\hat{\mathrm{x}}$, $\hat{\mathrm{f}}$ are estimations of $\mathrm{x}^{(0)}$, $\mathrm{f}^{(0)}$, respectively, here $\lambda > 0$ severs as a parameter of (2.1.2).

## 2.2 Recovery guarantee

We prove that (2.1.2) can recover $\mathrm{x}^{(0)}$ and $\mathrm{f}^{(0)}$ exactly with high probability provided that $\widetilde{\mathrm{A}}$, $\mathrm{x}^{(0)}$ and $\mathrm{f}^{(0)}$ satisfy certain conditions, which is stated formally in below theorem 2.1.

***Theorem 2.1*** *if n is prime, $\wedge$ and $\wedge\left(s_f^c\right)$ are random subsets of $[n]$, $\left|s_f^c\right| \geq \frac{32}{3}|s_x| ln\left(\frac{2|s_x|}{\varepsilon}\right)$ and $\left|s_f\right| \geq \frac{32}{3}|s_x| ln\left(\frac{2|s_x|}{\varepsilon}\right)$, where $0 < \varepsilon < \frac{1}{3}$ is a constant. Furthermore, if below (2.2.1) holds,*

[3] Here we define the sign of a complex variable $\mathrm{re}^{\mathrm{i}\theta}$ as $\mathrm{e}^{\mathrm{i}\theta}$, where $\mathrm{r} \geq 0$ denotes the absolute value of this variable, in particular, we define the sign of 0 as 0.

$$\rho_1\sqrt{|s_f|}+\rho_2\sqrt{|s_x|}\le\frac{1}{2}c\sqrt{|s_f^c|-|s_x|} \qquad (2.2.1)$$

*Where* $\rho_1=\sqrt{\frac{n}{m}}\lambda$ , $\rho_2=\sqrt{\frac{m}{|s_f^c|}}\sqrt{6}\left(\lambda^{-1}+\sqrt{2ln(2|s_x|/\varepsilon)}\right)+\sqrt{6}\left(1+\lambda\sqrt{2ln(2|s_x|/\varepsilon)}\right)\sqrt{\frac{n}{|s_f^c|}}$ , $0<c<1$ *is a constant, then with probability at least* $1-3\varepsilon$, *the solution of minimization (2.1.2) can recover* $x^{(0)}$ *and* $f^{(0)}$ *exactly.*

Assuming that $m=C_1 n$, $|s_f|=C_2 m$, where $C_1,C_2\in(0,1)$. In other words, the total number of measurements m takes a certain percentage of the dimension n, and the number of corrupted observation occupies a constant fraction of m, then one have the following conclusion according to theorem 2.1:

- When $\lambda=1$, theorem 2.1 shows that exact recovery by the solution of (2.1.2) is possible if $|s_x|\le C_3 m/\ln(|s_x|)$, provided that $C_3$ and $C_2$ are sufficiently small. Roughly speaking, when $\lambda=1$, theorem 2.1 states that if $|s_x|\le O(m/\ln(|s_x|))$, then (2.1.2) can recover $x^{(0)}$ and $f^{(0)}$ even when a constant fraction of the measurement vector b are arbitrarily corrupted, this bound is better than the optimal bound achieved by the BOS sensing matrix chap.12 [1], which is $O(m/\ln(n))$. However, condition (2.2.1) in theorem 2.1 prevents $|s_f|$ from becoming arbitrary close to m.
- When $\lambda=1/\sqrt{\ln(2n/\varepsilon)}$, theorem 2.1 shows that (2.1.2) can recover $x^{(0)}$ and $f^{(0)}$ if $|s_x|\le O(m/\ln(n))$ be sufficiently small, as it is indicated above, this upper bound for $|s_x|$ is also the asymptotically optimal bound when sensing matrix is partial BOS. Morover, it's worthy to mention that in this case, when n is sufficiently large, it allows that the number of corrupted observations $|s_f|$ grow arbitrarily close to the total number of measurement m, because in this case when $n\to\infty$, on has $\rho_1\to 0$ , this guarantees condition (2.2.1) in theorem 2.1 hold.

# Existing works and our contribution

Candes 2006 [4] shows that when $\widetilde{A}=F$ be the full Fourier basis matrix, then (2.1.2) can recover $x^{(0)}$ and $f^{(0)}$ exactly provided that $\|x^{(0)}\|_0+\|f^{(0)}\|_0\le O\left(\frac{n}{\sqrt{\ln(n)}}\right)$, the upper bound of $\|x^{(0)}\|_0$ in [4] can be larger than those proposed in theorem 2.1 in this paper by a factor of $\sqrt{\ln(n)}$, however, it disallows constant fraction of corruption in the measurement vector b, furthermore, there are restrictions on $x^{(0)}$ and $f^{(0)}$ as showed in sect. 2 of [4], e.g., the support of $x^{(0)}$ and $f^{(0)}$ are 2 independent random subsets of [n], and the non-zero elements in $f^{(0)}$ should obey circularly symmetric distribution in the complex plane.

Wright et. al [30] show that (2.1.2) can successfully recover $x^{(0)}$ and $f^{(0)}$ even if $f^{(0)}$ corrupted almost all entries in measurement vector b, provided $\|x^{(0)}\|_0$ is sub-linear smaller than m, which is far from the bound $|s_x|\le O(m/\ln(n))$ as we obtained in theorem 2.1, furthermore, the matrix $\widetilde{A}$ is an Gaussian i.i.d designed matrix, which is different from the partial Fourier basis matrix as we consider in this paper.

Li [2] and Nguyen [3] propose a reweighted $\ell_1$ minimization:

$$\text{minimize}_{x,f}\ \|x\|_1 + \theta\|f\|_1\ \text{s.t.}\ \widetilde{A}x + f = b \qquad (2.2.2)$$

After appropriately choosing $\theta$ depending on m and n, they also show that recovery of $x^{(0)}$ and $f^{(0)}$ by (2.6) is possible even the corruption ratio $\frac{\|f^{(0)}\|_0}{m}$ grows arbitrarily close to 1, provided that $m \geq O(\|x^{(0)}\|_0 \ln^2(n))$, which is asymptotically worse than the bound required in theorem 2.1 by a $\ln(n)$ factor.

Besides, all the literatures mentioned above [2-4] require certain restrictions on the $x^{(0)}$ and $f^{(0)}$ in their proofs, e.g. Wright [30] assumes that the support of $f^{(0)}$ distributes randomly and uniformly, and the signs of non-zero elements in $f^{(0)}$ takes $+1, -1$ uniformly and independently. In Nguyen [3] the author assumes the signs of $x^{(0)}$ on the support of $x^{(0)}$ are independently and uniformly to be +1 or -1, and the support of $f^{(0)}$ is a random subset of [m]. Li [2] assumes that the sign sequences of $x^{(0)}$ and $f^{(0)}$ on their supports are independent of each other, and each is a sequence of symmetric i.i.d $\pm 1$ variables.

In this paper, the support and sign of $x^{(0)}$ as well as the sign of $f^{(0)}$ can be arbitrary, the only restriction on $f^{(0)}$ is that $\wedge(s_f^c)$ should be a random subset of [n], to this end, one can only assume the support of $f^{(0)}$ be independent of $\wedge$ (which is the row indices set of $\widetilde{A}$ corresponding to the Fourier basis matrix F).

An unappealing restriction in theorem 2.1 is that it requires n to be prime, which is unlikely satisfied in real applications. To the best of our knowledge, whether the conclusion of theorem 2.1 holds for more general case where $n \in Z^+$ remains an open problem. However, according to the promising recovery results presented in section 4 when n is not prime, we believe that restricting n to be a prime is artificial and only for the convenience of the current proof in this paper.

## Stable recovery guarantee

Finally, although in theorem 2.1, it doesn't consider explicitly certain bounded, dense measurement noise, say $b = \widetilde{A}x^{(0)} + f^{(0)} + v$, where v is a m-dimensional vector denotes some kinds of measurement noise, e.g., v can be a vector of Gaussian noise. In fact, using the dual vector $h \in C^m$ we constructed in the next section, stable guarantee of the below quadratic programming is achievable according to theorem 3.33 of [1]:

$$\text{minimize}_{x,f}\ \|x\|_1 + \|f\|_1\ \text{s.t.}\ \|\lambda\widetilde{A}x + f - b\|_2 \leq \eta \qquad (2.2.3)$$

Where $\|z\|_2 \leq \eta$, which states that the recovery error of (2.2.3) is proportional to $\eta$ and $\sigma_k(x^{(0)})_1$, where $\sigma_k(\cdot)_1$ is defined below as in [1, 8]:

$\sigma_k(y)_1 = \min_x \|x - y\|_1, \text{s.t.} \|x\|_0 \leq k$ (2.2.4)

# 3. Proof road map of theorem 2.1

In this section, we provide a brief roadmap for the proof of theorem 2.1, where the necessary supporting lemma given in section A.1~A.2 which lead to a final proof of theorem 2.1 as stated in section A.3.

Firstly we introduce some notations that will be used in this section and in the rest of this paper.

*Notations.* In this paper, $a{:}b$ denotes the set $\{a, a+1, \ldots, b\}$ where a, b are 2 integers such that $a \geq b$. $M^*$ denotes the conjugate of M if M denotes a matrix with complex entries. $I_N$ denotes a $N \times N$ identity matrix, where N is a positive integer.

The starting point of the proof is the so called "dual certification" as presented in the following lemma 3.1:

***Lemma 3.1 (dual certification [39])*** *$x^{(0)}$ and $f^{(0)}$ is the unique solution of (2.1.2) if and only if the following condition holds:*

*There exists a vector $h \in C^m$ such that,*

$$\begin{cases} \lambda\tilde{A}^*(s_x,[m])h = \sigma_x, h(s_f) = \sigma_f \\ \left\|\lambda\tilde{A}^*(s_x^c,[m])h\right\|_\infty < 1, \left\|h(s_f^c)\right\|_\infty < 1 \end{cases} \quad (3.1)$$

*And matrix B=$[\lambda\tilde{A}([m], s_x), I_m([m], s_f)]$ is full rank*

*Where $\tilde{A}^*$ denotes the conjugate matrix of $\tilde{A}$.*

*Proof: this follows from a direct application of theorem 4 in [39].* ■

According to [30], we have the below lemma 3.2.

***Lemma 3.2*** *([30]) suppose we have a $|s_x^c| + |s_f^c|$ dimensional vector q satisfies,*

$$\begin{bmatrix} \lambda\tilde{A}^*(s_x^c, s_f^c) & I_{|s_x^c|} \\ \lambda\tilde{A}^*(s_x, s_f^c) & 0 \end{bmatrix} q = \begin{bmatrix} -\lambda\tilde{A}^*(s_x^c, s_f)\sigma_f \\ \sigma_x - \lambda\tilde{A}^*(s_x, s_f)\sigma_f \end{bmatrix} \quad (3.2)$$

*And $\|q\|_\infty < 1$, where $I_{|s_x^c|}$ denotes a $|s_x^c| \times |s_x^c|$ identity matrix. then there exists a m-dimensional vector h as defined in (3.3) satisfies (3.1).*

$h(s_f) = \sigma_f,\ h(s_f^c) = q(1{:}|s_f^c|)$ (3.3)

Proof:

According to (3.3) and (3.2), one has,

$$\begin{cases} \lambda\tilde{A}^*(s_x^c, s_f^c)h(s_f^c) + q(|s_f^c|+1{:}|s_f^c|+|s_x^c|) = -\lambda\tilde{A}^*(s_x^c, s_f)h(s_f) \\ \lambda\tilde{A}^*(s_x, s_f^c)h(s_f^c) = \sigma_x - \lambda\tilde{A}^*(s_x, s_f)h(s_f) \end{cases}$$

Which implies that:

$$\begin{cases} q(|s_f^c|+1{:}|s_f^c|+|s_x^c|) = -\lambda\tilde{A}^*(s_x^c,[m])h \\ \lambda\tilde{A}^*(s_x,[m])h = \sigma_x \end{cases} \quad (3.4)$$

Combining (3.4), (3.2) and the fact that $\|q\|_\infty < 1$ proves the conclusion of the lemma. ■

By lemma 3.1~lemma 3.2, we conclude that if we can prove that matrix B in lemma 3.1 is full rank and then find a q satisfies (3.2) and $\|q\|_\infty < 1$, then we can prove that $x^{(0)}$ and $f^{(0)}$ can be recovered by the solution of (2.1) exactly, lemma 3.3 below shows a sufficient condition for the existence of such q, before presenting the lemma, we define the soft operator [30] which is used in the remaining of this section.

***Definition 3.1 (soft thresholding operator [30])*** *if* $x \in C$ *denotes a complex variable,* $\mathcal{S}(x) = \begin{cases} x, if\ |x| \geq 1/2 \\ 0, otherwise \end{cases}$ *defines the soft thresholding value of* $x$*, here* $\mathcal{S}(\cdot)$ *denotes the soft thresholding operator, similarly, if* $x \in C^n, n \geq 1$ *denotes a n-dimensional complex vector, then* $y = \mathcal{S}(x)$ *denotes a n-dimensional vector with* $y(i) = \mathcal{S}(x(i)), 1 \leq i \leq n$*.*

***Lemma 3.3*** *(lemma 2 in [30]) Consider an underdetermined system* $\Phi q = w$*. let* $P_\Phi$ *denotes the projection operator on to* $range(\Phi^*)$*, and let* $\xi_k(\Phi)$ *denotes the norm of* $P_\Phi$*, restricted to sparse vectors:*

$$\xi_k(\Phi) := sup_{\|x\|_0 \leq k, \|x\|_2 \leq 1} \|P_\Phi x\|_0 \qquad (3.5)$$

*Suppose* $\xi_k(\Phi) < 1$ *and there exists a solution* $q_0$ *satisfies,*

$$\Phi q_0 = w,\ \|q_0\|_2 + \frac{\xi_k(\Phi)}{1-\xi_k(\Phi)} \|\mathcal{S}[q_0]\|_2 \leq \sqrt{k}/2 \qquad (3.6)$$

*Where* $\mathcal{S}$ *is the soft-thresholding operator with threshold 1/2, then there exists a solution q to system* $\Phi q = w$ *satisfies* $\|q\|_\infty < 1/2$*.*

Since $\|\mathcal{S}[q_0]\|_2 \leq \|q_0\|_2$, a sufficient condition of $q_0$ satisfies (3.6) is $q_0$ satisfies below (3.7):

$$\Phi q_0 = w,\ \left(1 + \frac{\xi_k(\Phi)}{1-\xi_k(\Phi)}\right) \|q_0\|_2 \leq \sqrt{k}/2 \qquad (3.7)$$

Let

$$\Phi = \begin{bmatrix} \lambda \tilde{A}^*(s_x^c, s_f^c) & I_{|s_x^c|} \\ \lambda \tilde{A}^*(s_x, s_f^c) & 0 \end{bmatrix},\ w = \begin{bmatrix} -\lambda \tilde{A}^*(s_x^c, s_f)\sigma_f \\ \sigma_x - \lambda \tilde{A}^*(s_x, s_f)\sigma_f \end{bmatrix} \qquad (3.8)$$

By the discussion above, the recovery guarantee of (2.1.2) is achieved in 3 steps:

Firstly, the proof of $\xi_k(\Phi) < 1$ when $k = \sqrt{|s_f^c| - |s_x|}$ is provided in appendix A.1, and the searching for a viable $q_0$ satisfying (3.7) are detailed in appendix A.2. Secondly, the proof of B in lemma 3.1 is full rank is provided in appendix A.3. Finally combining results in appendix A.1~A.3 leads to a natural proof of theorem 2.1, which is provided in appendix A.4.

# 4. Experiments

In this section, we provide numerical experiments based on synthesis data as well as real image data to verify the conclusion of theorem 2.1, although the requirement of n is a prime in theorem 2.1 is necessary in our proof, but it is not necessary satisfied in real applications, to understanding the performance of (2.1.2) when this assumption is violated, we also show the experiment results when n is not prime in each sections. For simplicity, we set $\lambda = 1$ in algorithm (2.1.2) within all our experiments, when $\lambda = 1/\sqrt{\ln(n/\varepsilon)}$, the recovery results are similar which are not presented in this paper for saving space.

## 4.1. Experiment based on synthesis data

In this section, we provide extensive simulations to illustrate the probability of the $\ell_1$-minimization (2.1.2) in recovering data $x^{(0)}$ and $f^{(0)}$.

The setting for $x^{(0)}$ and $f^{(0)}$ are according to theorem 2.1 as described below: $\|x^{(0)}\|_0 = 0.2n/\ln(0.2n)$, which is asymptotically the same as suggested in theorem 2.1, $\widetilde{A} = \frac{1}{\sqrt{m}}F(\wedge, [n])$, where $\wedge$ is selected randomly and uniformly in [n], to ensure $\wedge\,(s_f^c)$ being a random subset of $[n]$, $s_f$ is chosen independently of $\wedge$.

Furthermore, to violate the random distribution assumptions on the supports and signs of $x^{(0)}$ and $f^{(0)}$ which are necessary in existing literatures [2-4] etc, but not necessary in theorem 2.1 in this paper, we set $s_x = \{j, j+1, \dots, j+|s_x|\}$, where j is selected randomly and uniformly in interval $[1, n-|s_x|]$, and without loose of generality, all entries in $x^{(0)}(s_x)$ and $f^{(0)}(s_f)$ are set to be positive. The norm of corrupted noise is set significantly larger than those of $x^{(0)}$, i.e, $\|f^{(0)}\|_2 \geq 100\|x^{(0)}\|_2$.

Random vectors $x^{(0)}$ and $f^{(0)}$ as described above are generated of varying parameters $\vartheta_m$, $\vartheta_f$ and $n$ to test the recoverability of algorithm (2.1.2). Where $\vartheta_m := \frac{|\wedge|}{n}$ denotes the measurement rate, $\vartheta_f := \frac{|s_f|}{m}$ denotes the corruption rate, and n denotes the dimension of the signal $x^{(0)}$.

The algorithm (2.1.2) is declared to succeed in each run if the relative recovery error (RRE) defined below is less than $10^{-8}$.

$$\mathrm{RRE} = \frac{\left\|\begin{bmatrix}\hat{x}-x^{(0)}\\ \hat{f}-f^{(0)}\end{bmatrix}\right\|_2}{\left\|\begin{bmatrix}x^{(0)}\\ f^{(0)}\end{bmatrix}\right\|_2} \qquad (4.1.1)$$

More specifically, different parameters $\vartheta_m$, $\vartheta_f$ and $n$ are set within a range as described below: $\vartheta_m \in \{0.1, 0.2, \dots, 1\}$ takes 10 different values, $\vartheta_f \in \{0.05, 0.15, 0.25, 0.35\}$ takes 4 different values, n takes different possible values within 2 different sets: in the first scenario, $n$ takes 20 prime values within the interval $[128, 512]$ while in the second scenario, n takes 20 non-prime integers which are chosen randomly and uniformly from interval $[128, 256]$.

The succeed rate of recovery of algorithm (2.1.2) in these 2 different scenarios are summarized in the 2 heat-maps of figure (4.1) respectively. In figure (4.1), the vertical axis in the heat-map indicates different values of $\vartheta_m$(the measurement rate), while the horizontal axis indicates different values of $\vartheta_f$ (the corruption rate) .

For each fixed parameters pair $(\vartheta_m, \vartheta_f, n)$, 25 independent runs of minimization (2.1.2) are performed in order to calculate the succeed rate, thus, each element in the heat maps of figure 4.1 is achieved through $20 \times 25 = 500$ independent runs of (2.1.2). e.g., the element locates on the first row, first column of the left heat map in figure 4.1 indicates the average succeed rate (the probability of RRE $< 10^{-8}$ occurs) of 20 different parameter pairs : $\{(\vartheta_m, \vartheta_f, n) | \vartheta_m = 0.1, \vartheta_f = 0.05, \text{n is prime belong to } [128,512]\}$.

For each fixed parameter pair $(\vartheta_m, \vartheta_f, n)$ we perform 25 independent runs and hence we totally performed $20 \times 25 = 500$ independent runs before we obtaining the average succeed rate value in this particular location of the heat-map. Elements in the right heat-map can be interpreted similarly.

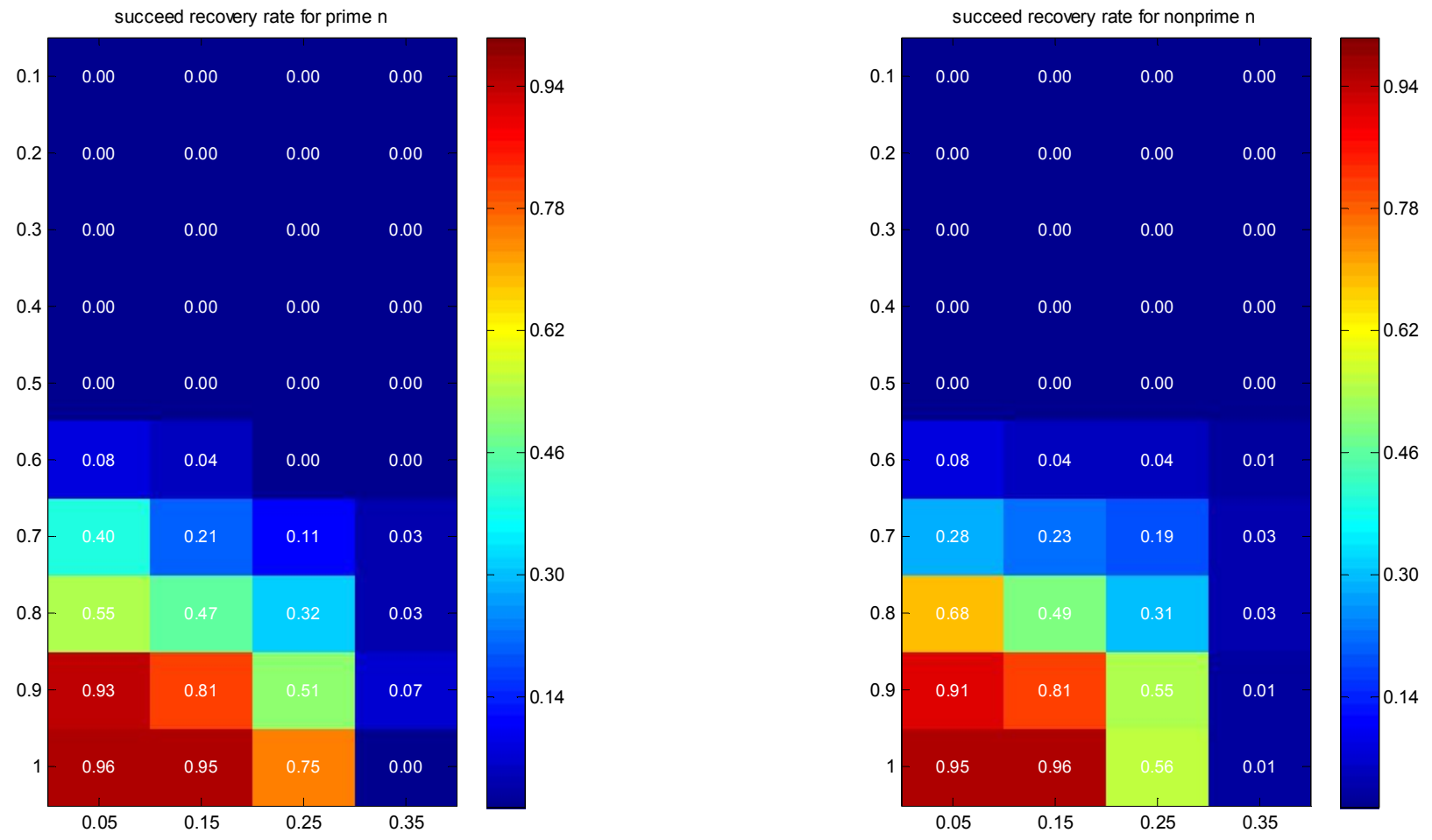

Figure (4.1) the recovery performance of (2.1.2) based on synthesis data. The left heat-map indicates the average recovery rate with different measurement rate $\vartheta_m$ (indicated on the vertical axis) and different corruption rate $\vartheta_f$ (indicated on the horizontal axis) when the signal dimension n is prime value in interval [128, 512]. The right heat-map is interpreted similarly when n is not prime integer in interval [128, 512].

As showed in figure (4.1), when the measurement rate $\vartheta_m \geq 0.9$ and the corruption rate $\vartheta_f \leq 0.15$, then the successful recovery rate of (2.1.2) is larger than 81% in most case, which validates the conclusion of theorem 2.1, also note that the recovery performance of (2.1.2) is not severely degraded when the assumption of n is prime is violated, as we can see in figure (4.1), there are only slightly difference between the left and the right heat map which is caused by the difference between the distribution of primes and non-prime integers in the interval [128, 512].

## 4.2 Experiments based on the real data

In this section, the signal to be recovered $x^{(0)}$ is chosen from the real world data, more specifically, data in $x^{(0)}$ is from an image patch selected randomly from the natural images in the BSDS-500 database , then elements of the image patch are stacked together to form $x^{(0)}$. 10 typical image patches selected in this manner are showed in figure (4.2.1). Without loss of generality, the size of the image path is set different values as $8 \times 8$, $16 \times 16$ and $32 \times 32$, the corresponding dimension of $x^{(0)}$ are set differently as 64, 256 and 1024.

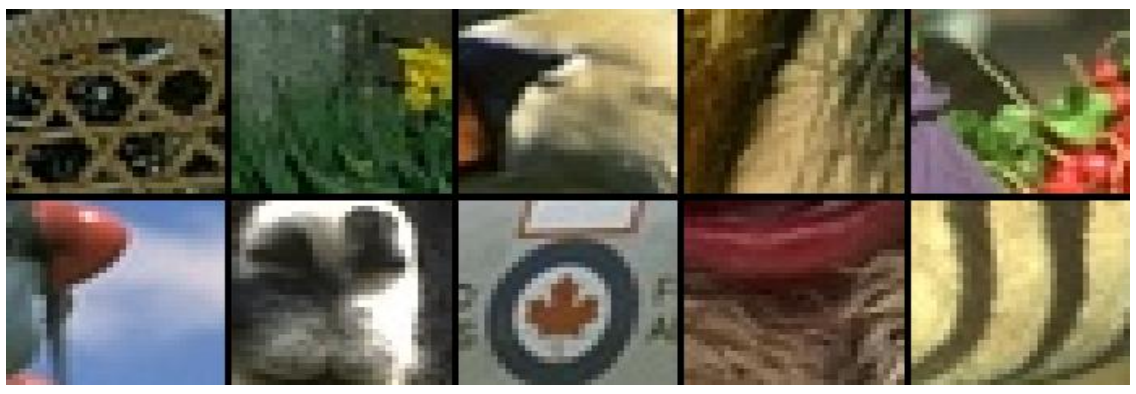

Figure (4.2.1) 10 sample $32 \times 32$ image patches randomly grabbed from the BSDS-500 data base

To measure the recovery performance of minimization (2.1.2) in this section, instead of recover $x^{(0)}$, we recover the Fourier transform of $x^{(0)}$(which is denoted as $\check{x}^{(0)}$) based on partial measurement of $x^{(0)}$, let $b = x^{(0)}(\wedge) + f^{(0)}$, where $\wedge$ is a random subset of [n], to ensure $\wedge\,(s_f^c)$ also be a random subset of [n], we set $s_f$ be independent of $\wedge$, then the minimization (2.1.2) to recover $\check{x}^{(0)}$ is wrote as:

$$\text{minimize } \|x\|_1 + \|f\|_1 \text{ s.t.} \widetilde{A}x + f = \widetilde{A}\check{x}^{(0)} + f^{(0)} \qquad (4.2.1)$$

Where $\widetilde{A} = \frac{1}{\sqrt{\wedge_m}} F^*(\wedge_m, [n])$ denotes the partial matrix of the conjugate Fourier basis with normalized columns, it is easily to verify that theorem 2.1 also applies to (4.2.1) since lemma A.1.1 applies when the measurement matrix is $F^*$.

The norm of corrupted noise is set significantly larger than $\|x^{(0)}\|_2$ as in the previous section, i.e, $\|f^{(0)}\|_2 \geq 100\|x^{(0)}\|_2$, finally, the recovery performance is measured by the signal relative recovery error (SRRE):

$$SRRE = \frac{\|\hat{x} - \check{x}^{(0)}\|_2}{\|\check{x}^{(0)}\|_2} \qquad (4.2.2)$$

Figure (4.2.1) displays the process of the recovery of a typical image patch mentioned above.

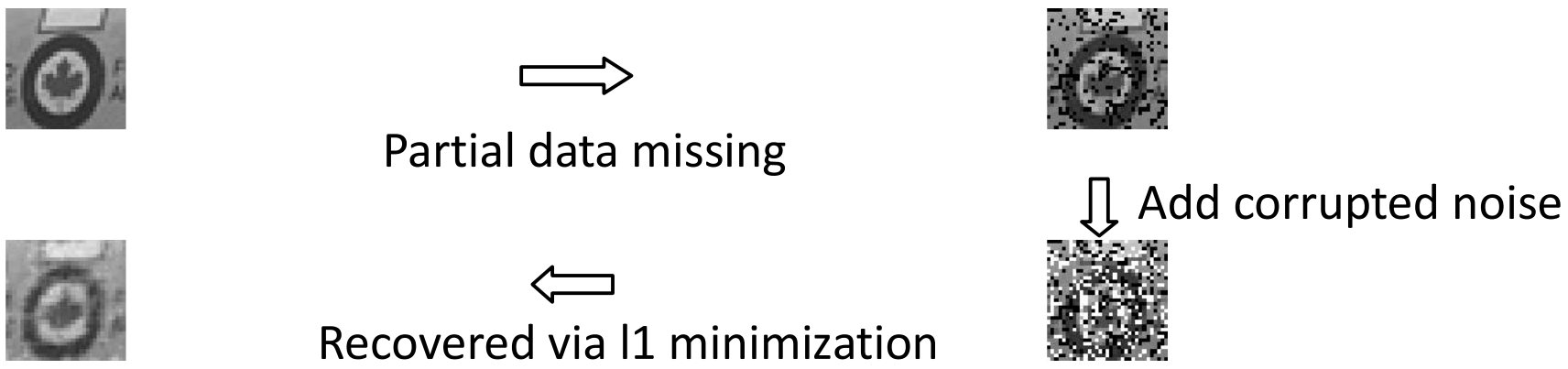

Figure (4.2.1) demonstration of the data recovery based on the real image data, the upper left image shows a typical $32 \times 32$ image patch randomly selected from BSDS 500 data base, the upper-right image shows the original image patch with 20% missing data (showed as the dark pixels), the lower-right image shows the upper-right image with 25% measurement are corrupted (showed as the bright pixels), finally the lower-left image shows the recovered image via the $\ell_1$ minimization (2.1.2) using the lower-right noisy image as input, in this example, the recovery relative error is 20.04%.

Unlike in the previous section where the signal to be recovered is designed to be sparse, $\check{x}^{(0)}$ here can only be treated as approximately sparse, for convenience, we let $\frac{\sigma_k(\check{x}^{(0)})_1}{\|\check{x}^{(0)}\|_2}$ be a k-sparse indicator of a non-zero vector $\check{x}^{(0)}$ which reflects how good $\check{x}^{(0)}$ can be approximated by a k-sparse vector, where the meaning of $\sigma_k(\check{x}^{(0)})_1$ is defined by (2.2.4).

It is easily to see that k-sparse indicator of a vector equals to 0 if and only if the cardinality of the vector is less than or equal to k. Moreover, the k-sparse indicator $\frac{\sigma_k(\check{x}^{(0)})_1}{\|\check{x}^{(0)}\|_2}$ also reflects a theoretical upper-bound of the relative recovery error on the convex programming (2.2.3). Figure (4.2.2) shows the k-sparse indicator curves of 3 different types of signal: 1. The signal is a Gaussian random vector; 2. The signal is a sparse random vector as described in the previous section; 3. The signal is $\check{x}^{(0)}$ as described above in this section.

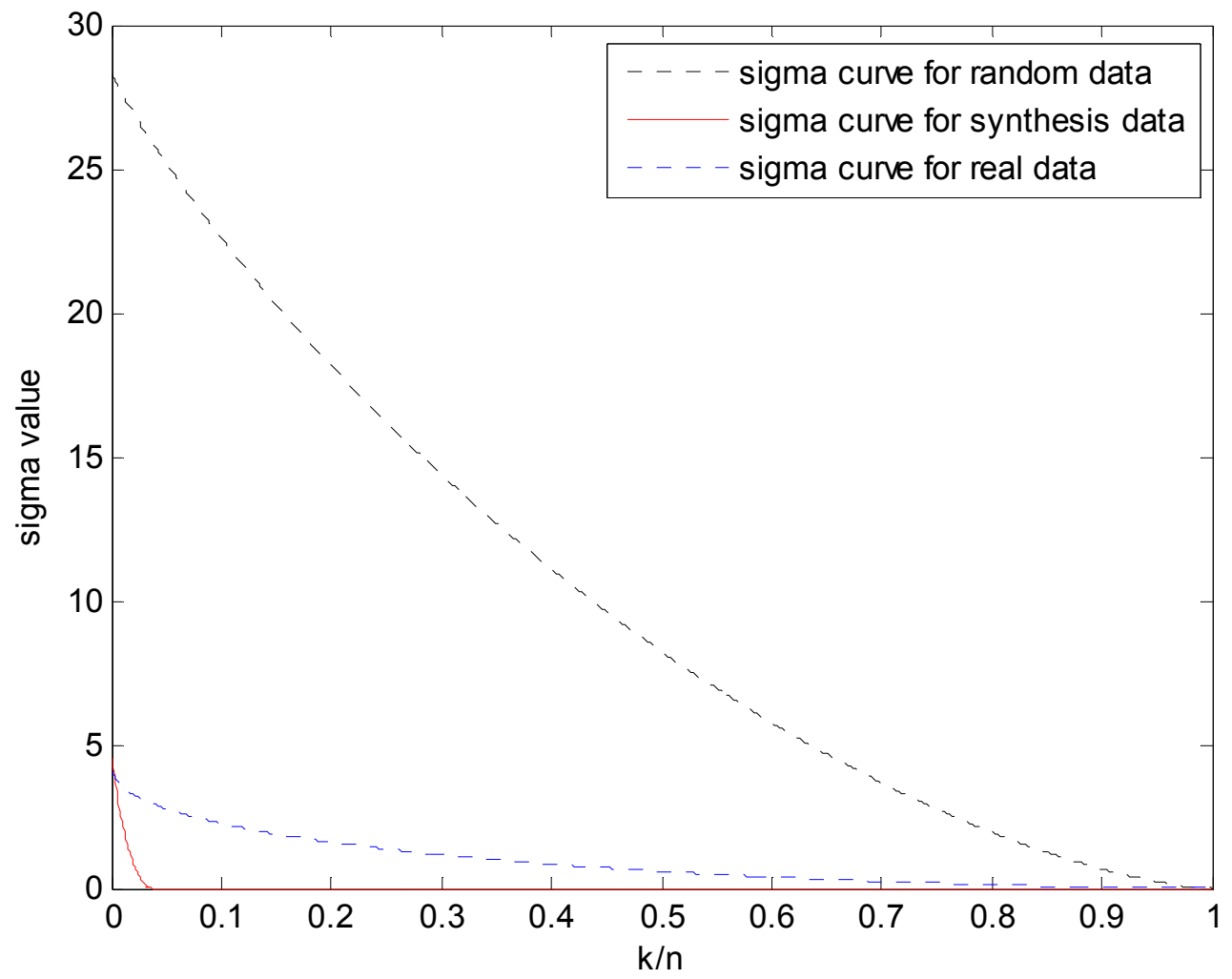

Figure (4.2.2) the average value of $\frac{\sigma_k(y)_1}{\|y\|_2}$ when y is the signal selected in 3 different cases. The horizontal axis indicates $\frac{k}{n}$ where n denotes the dimension of y, and $1 \leq k \leq n$ be positive integers, the vertical axis denotes the average value of $\frac{\sigma_k(y)_1}{\|y\|_2}$. i.e., the black dashed curve , the blue dashed cure and the red soild curve indicate the average value of $\frac{\sigma_k(y)_1}{\|y\|_2}$ of 200 random signals, 200 synthesis signal $x^{(0)}$ as defined in section 4.1 and 200 Fourier transform of $32 \times 32$ image patch signal defined in this section, respectively.

In figure (4.2.2), we observe that the k-sparse indicator curve of $\check{x}^{(0)}$ defined in this section (the dashed blue curve) decreases more rapidly than the curve of the random signal (the dashed dark curve) which suggests that $\check{x}^{(0)}$ we used in this section can be treated approximately a k-sparse vector if k is sufficiently large. Also notice that we have $\frac{\sigma_k(\check{x}^{(0)})_1}{\|\check{x}^{(0)}\|_2} > 0$ whenever $k < n$, this suggests that $\check{x}^{(0)}$ can be represented as a k-sparse vector plus a bounded non-zero vector.

The simulation results of this section is summarized in the heat-maps of figure (4.2.3), where each element in a heat-map indicates the average SRRE value (obtained by averaging the SRRE of 200 image patch selected randomly from BSDS-500) for a particular $\vartheta_m$, $\vartheta_e$ pair value.

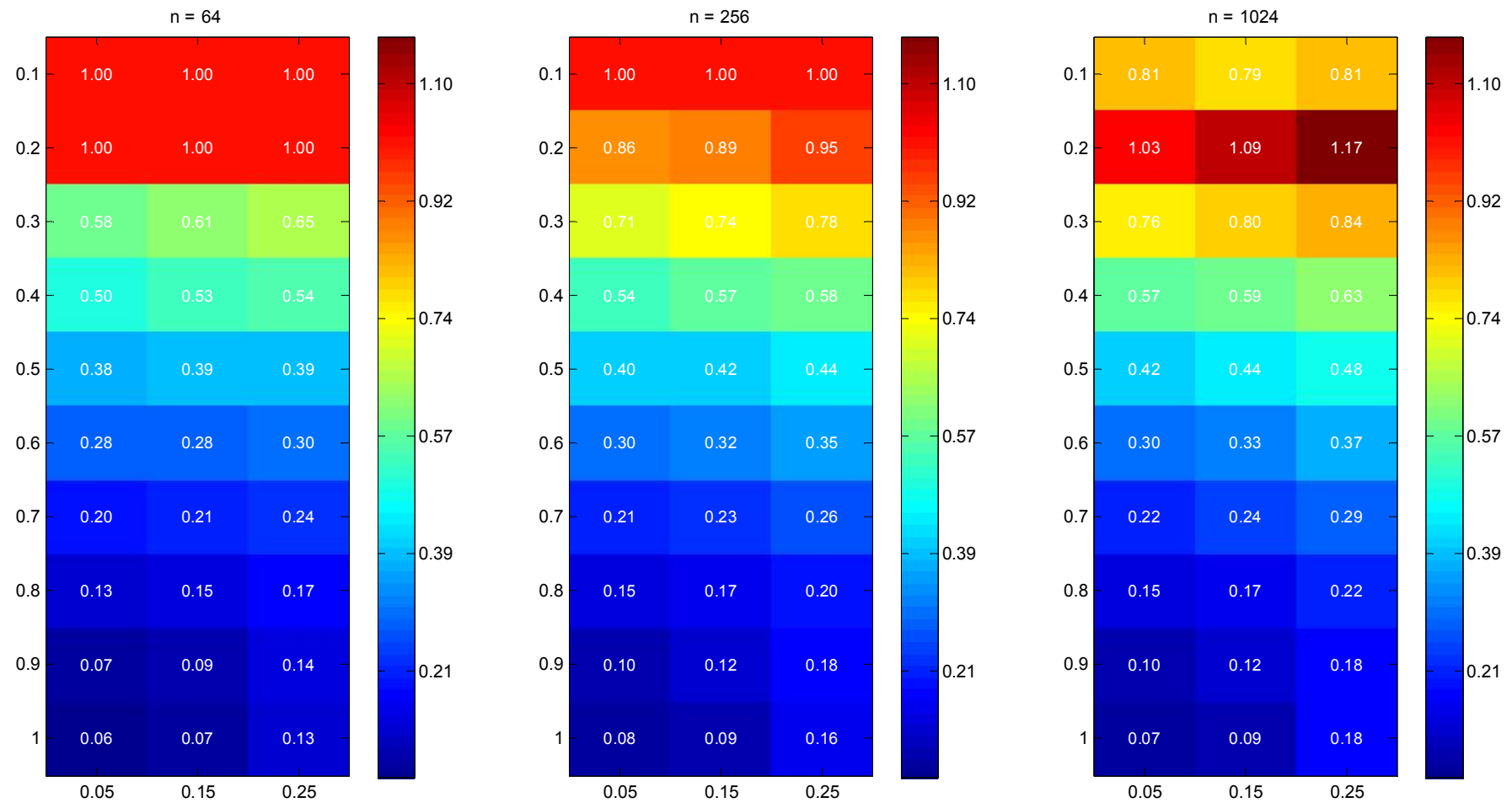

Figure (4.2.3) performance of (2.1.2) based on the real world data. Each element in the heat-maps indicates the average SRRE value obtained by 200 independent runs based on the real world signal $x^{(0)}$, with a particular parameters pair $(n, \vartheta_m, \vartheta_f)$. The vertical axis of the heat-maps indicates different value of $\vartheta_m$ (the measurement rate), the horizontal axis of the heat-maps indicates different value of $\vartheta_f$ (the corruption rate), and the title of the heat-maps indicates the dimensions of the input signal $x^{(0)}$. E.g., the element on the first row, first column of the left-most heat-map indicates the average SRRE based on 200 $8 \times 8$ different image patches randomly selected from BSDS 500, with $\vartheta_m = 0.1$ and $\vartheta_f = 0.05$

It can be seem from figure (4.2.3) that minimization (2.1.2) works well on recovering real image data when the measurement ratio $\vartheta_m$ is large enough and the corruption ratio $\vartheta_f$ is small enough (e.g., the average relatively recovery error of $x^{(0)}$ is less than 12% when $\vartheta_m \geq 0.9$ and $\vartheta_f \leq 15\%$), and the SRRE value displayed in figure (4.2.2) gradually decrease as $\vartheta_m$ decreases and $\vartheta_f$ increases, this verify the conclusion of theorem 2.1 even when n (the length of signal) is not prime.

Notice that the recovery result in figure (4.2.2) is not as good as those in figure (4.1.1), where in figure (4.1.1) one obtain accurate recovery results with certain probability, but in figure (4.2.2), the SRRE value is always larger than a positive constant, this can be explained by the fact that $\check{x}^{(0)}$ is only approximately sparse as suggested in the k-sparse indicator curve in figure (4.2.2).

# 5. Conclusion and future works

In summary, we prove that exact recovery of signal is possible from linear measurements $b = Ax^{(0)} + f^{(0)}$ even when a constant fraction of measurement b is arbitrarily corrupted by $f^{(0)}$, unlike in the existing literatures [2, 3, 30], the assumptions on the signs and supports of the $f^{(0)}$ and $x^{(0)}$ are kept to be minimal: we only require the support of $f^{(0)}$ be chosen such that $\wedge\left(s_f^c\right)$ is a random subset of [m]. Extensive experiments based on the synthesis data as well as the real world data faithfully verify the conclusion of theorem 2.1 even when n is not prime, furthermore, we've also observed similar experimental results when the sensing matrix $A$ is chosen as partial Bounded Orthogonal System [1] other than partial Fourier basis. These observations drive us to believe that similar conclusion as in theorem 2.1 might hold for general integer n and when A is belong to a broader family—the partial BOS, and the generalization of theorem 2.1 in such kinds are left in future works.

# Acknowledgement

The author wish to thank his wife Haixia Yu for her mental as well as material supports over the past 8 years.

# Appendix A—supporting lemmas for theorem 2.1

For convenience, firstly we introduce 2 definitions which would be used later in this section.

***Definition A.1*** *(Bounded Orthonormal System chap.12 [1]) We call* $\Psi = \{\psi_1, \dots, \psi_N\}$ *a bounded orthonormal system (BOS) with constant K if it satisfies (A.1):*

$$\begin{cases} \psi_j^* \psi_k = \begin{cases} 0, if\ j \neq k \\ 1, if\ j = k \end{cases} \\ \left\|\psi_j\right\|_\infty \leq K \end{cases} \text{for all } j,k \in [N] \qquad (A.1)$$

***Definition A.2*** *(Steinhaus sequence, [1]) A complex random variable which is uniformly distributed on the torus* $T = \{z \in C, |z| = 1\}$ *is called a Steinhaus variable. A sequence* $\epsilon = (\epsilon_1, \dots, \epsilon_N)$ *of independent Steinhaus variables is called a Steinhaus sequence.*

## A.1 Proving $\xi_k(\Phi) < 1$

***Lemma A.1.1*** *([40]) let f be a n-dimensional non-zero complex vector,* $\check{f}$ *be the Fourier transform of f, suppose n is prime, then,*

$$\|f\|_0 + \|\check{f}\|_0 \geq n + 1 \qquad (A.1.1)$$

***Theorem A.1.1*** *if* $k \leq \left|s_f^c\right| - |s_x|$ *and n is prime then,* $\xi_k(\Phi) < 1$ *holds.*

Proof:

Event $\xi_k(\Phi) < 1$ is equivalent to the event that a k-sparse $(|s_x^c| + |s_f^c|)$-dimensional vector y cannot be represented by the column space of $\Phi^* = \begin{bmatrix} \lambda\tilde{A}(s_f^c, s_x^c) & \lambda\tilde{A}(s_f^c, s_x) \\ I_{|s_x^c|} & 0 \end{bmatrix}$, suppose the inverse is true, i.e., let z be a n-dimensional non-zero vector satisfies,

$$\begin{bmatrix} \lambda\tilde{A}(s_f^c, s_x^c) & \lambda\tilde{A}(s_f^c, s_x) \\ I_{|s_x^c|} & 0 \end{bmatrix} z = y, \text{ such that } \|y\|_0 \le k \qquad \text{(A.1.2)}$$

Let $\check{z}$ denotes the Fourier transform of z, since $\lambda$ is a positive constant, (A.1.2) implies that,

$$\|y\|_0 = \|\check{z}(\wedge(s_f^c))\|_0 + \|z(1{:}|s_x^c|)\|_0 \ge \|\check{z}(\wedge(s_f^c))\|_0 + \|z\|_0 - |s_x| \ge \|\check{z}\|_0 + \|z\|_0 - \|\check{z}(\wedge^c)\|_0 - \|\check{z}(\wedge(s_f))\|_0 - |s_x| > |s_f^c| - |s_x| \qquad \text{(A.1.3)}$$

Which leads to a contradiction since $\|y\|_0 \le k \le |s_f^c| - |s_x|$, where the last inequality of (A.1.3) we use the fact $\|\check{z}\|_0 + \|z\|_0 \ge n + 1$as indicated by lemma A.1.1.■

## A.2 Finding a viable $q_0$ satisfies (3.7)

To find a viable $q_0$ satisfies (3.7), firstly we find a $h_0$ as below:

$$\begin{cases} h_0(s_f) = \sigma_f \\ h_0(s_f^c) = \lambda^{-1}\tilde{A}^{\dagger}(s_f^c, s_x)\sigma_x' \end{cases} \qquad \text{(A.2.1)}$$

Where $\tilde{A}^{\dagger}(s_f^c, s_x) = \tilde{A}(s_f^c, s_x)\left(\tilde{A}^*(s_x, s_f^c)\tilde{A}(s_f^c, s_x)\right)^{-1}$ represents the Peron-Moore inverse of $\tilde{A}(s_f^c, s_x)$, $\sigma_x' = \sigma_x - \lambda\tilde{A}^*(s_x, s_f)\sigma_f$. Then we obtain $q_0$ as in (A.2.2), which is stated below,

$$\begin{cases} q_0(1{:}|s_f^c|) = h_0(s_f^c) \\ -q_0(|s_f^c| + 1{:}|s_f^c| + |s_x^c|) = \lambda\tilde{A}^*(s_x^c, [m])h_0 \end{cases} \qquad \text{(A.2.2)}$$

It is easy to see that $q_0$ as defined in (A.2.2) satisfies $\Phi q_0 = w$, where $\Phi, w$ are defined as in (3.8). The remaining part is to bound $\|q_0\|_2$.

We'll firstly bound $\|h_0\|_2$, and then bound $\|q_0\|_2$ according to (A.2.2).

***Lemma A.2.1 (theorem 12.12 of [1])*** *Let $A \in C^{m\times n}$ be random sampling matrix associated to a Bounded Orthogonal System (BOS) with constant $K \ge 1$. Let $s \subset [n]$ be an indices set. Then, for $\delta \in (0,1)$, the normalized matrix $\tilde{A} = \frac{1}{\sqrt{m}}A$ satisfies $\left\|\tilde{A}^*([m], s)\tilde{A}([m], s) - I_{|s|}\right\|_{2\to 2} \le \delta$, with probability at least $1 - 2|s|exp\left(-\frac{3m\delta^2}{8K^2|s|}\right)$.*

***Lemma A.2.2*** *if $\wedge$ and $\wedge(s_f)$ are random subsets of $[n]$, then,*

$$prob\left(\left\|\tilde{A}^*(s_x, s_f)\right\|_{2\to 2} \le \sqrt{\frac{|s_f|}{m}}\sqrt{\frac{3}{2}}, \left\|\tilde{A}^{\dagger}(s_f^c, s_x)\right\|_{2\to 2} \le \sqrt{\frac{m}{|s_f^c|}}\sqrt{6}\right) \ge 1 - 2\varepsilon \qquad \text{(A.2.3)}$$

*holds for any $\varepsilon \in (0, \frac{1}{2})$, provided that $|s_f^c| \ge \frac{32}{3}|s_x| ln\left(\frac{2|s_x|}{\varepsilon}\right)$ and $|s_f| \ge \frac{32}{3}|s_x| ln\left(\frac{2|s_x|}{\varepsilon}\right)$.*

Proof:

Substituting the constant K=1 and letting $\delta = 1/2$ into lemma A.2.1, one has,

$$\text{prob}\left(\left\|\frac{m}{|s_f^c|}\tilde{A}^*(s_x, s_f^c)\tilde{A}(s_f^c, s_x) - I_{|s_x|}\right\|_{2\to 2} \le \frac{1}{2}\right) \ge 1 - \varepsilon \qquad \text{(A.2.4)}$$

provided that,

$$|s_f^c| \ge \frac{32}{3}|s_x| \ln\left(\frac{2|s_x|}{\varepsilon}\right) \qquad \text{(A.2.5)}$$

Holds, similarly,

$$\mathrm{prob}\left(\left\|\frac{m}{|s_f|}\widetilde{A}^*(s_x,s_f)\widetilde{A}(s_f,s_x)-I_{|s_x|}\right\|_2\le\frac{1}{2}\right)\ge 1-\varepsilon \qquad \text{(A.2.6)}$$

provided that,

$$|s_f|\ge\frac{32}{3}|s_x|\ln\left(\frac{2|s_x|}{\varepsilon}\right) \qquad \text{(A.2.7)}$$

Consequently, suppose (A.2.5)and (A.2.7) holds, one have,

$$\mathrm{prob}\left(\sqrt{\frac{m}{|s_f|}}\left\|\widetilde{A}^*(s_x,s_f)\right\|_{2\to 2}\le\sqrt{\frac{3}{2}}\right)\ge 1-\varepsilon \quad \text{and} \quad \mathrm{prob}\left(\left\|\widetilde{A}^\dagger(s_f^c,s_x)\right\|_{2\to 2}\le\sqrt{\frac{m}{|s_f^c|}}\sqrt{6}\right)\ge 1-\varepsilon$$

holds, applying the union bound proves the conclusion of the lemma.■

***Lemma A.2.3*** *(Hoeffding-type inequality for Steinhaus sums, corollary 8.10 in [1]) Let* $a\in C^M$ *and* $\epsilon=(\epsilon_1,\dots,\epsilon_M)$ *be a Steinhaus sequence. For any* $0<\gamma<1$,

$$prob\left(\left|\sum_{j=1}^M\epsilon_j a(j)\right|\ge u\|a\|_2\right)\le\frac{1}{1-\gamma}exp(-\gamma u^2)\ \text{for all } u>0 \qquad \text{(A.2.8)}$$

***Lemma A.2.4*** *if* $\wedge(s_f)$ *is a random subset of [n], then,*

$$prob\left(\left\|\tilde{A}^*(s_x,s_f)\sigma_f\right\|_2\le\sqrt{2|s_x|ln(2|s_x|/\varepsilon)}\right)\ge 1-\varepsilon \qquad \text{(A.2.9)}$$

*for any* $0<\varepsilon<1$.

Proof:

Let $\alpha=\widetilde{A}^*(s_x,s_f)\sigma_f$, then $\alpha(j)=\widetilde{A}^*(s_x(j),s_f)\sigma_f, 1\le j\le|s_x|$, since $\wedge(s_f)$ is a random subset of [n], elements in the row vector $\sqrt{m}\widetilde{A}^*(s_x(j),s_f), 1\le j\le|s_x|$ form a Steinhaus sequence, therefore according to lemma A.2.3 and letting $\gamma=\frac{1}{2}$, $u=\sqrt{2\ln(2|s_x|/\varepsilon)}$, one has,

$$\mathrm{prob}\left(|\alpha(j)|\le\frac{\sqrt{|s_f|}}{\sqrt{m}}\sqrt{2\ln(2|s_x|/\varepsilon)}\right)\ge 1-2\exp\left(-\ln(2|s_x|/\varepsilon)\right)=1-\frac{\varepsilon}{|s_x|}\quad ,\qquad 1\le j\le|s_x|$$

(A.2.10)

then applying a union bound on (A.2.10) and using the fact $\frac{\sqrt{|s_f|}}{\sqrt{m}}\le 1$ prove (A.2.9). ■

***Lemma A.2.5*** *if conditions in lemma A.2.4 holds, then,*

$$prob\left(\|\sigma_x'\|_2\le\left(1+\lambda\sqrt{2ln(2|s_x|/\varepsilon)}\right)\sqrt{|s_x|}\right)\ge 1-\varepsilon \qquad \text{(A.2.11)}$$

*for any* $0<\varepsilon<1$.

Proof:

Since $\sigma_x'=\sigma_x-\lambda\widetilde{A}^*(s_x,s_f)\sigma_f$, by triangle inequality: $\|\sigma_x'\|_2\le\|\sigma_x\|_2+\lambda\left\|\widetilde{A}^*(s_x,s_f)\sigma_f\right\|_2$, then the conclusion of the lemma follows immediately from (A.2.9). ■

***Lemma A.2.6*** *(bounding* $\left\|h_0(s_f^c)\right\|_2$*) If conditions in lemma A.2.2 hold,* $h_0$ *is defined in (A.2.1), then,*

$$prob\left(\left\|h_0(s_f^c)\right\|_2\le\sqrt{\frac{m}{|s_f^c|}}\sqrt{6}\left(\lambda^{-1}+\sqrt{2ln(2|s_x|/\varepsilon)}\right)\sqrt{|s_x|}\right)\ge 1-3\varepsilon \qquad \text{(A.2.12)}$$

Proof:

According to (A.2.1), $h_0(s_f^c)=\lambda^{-1}\widetilde{A}^\dagger(s_f^c,s_x)\sigma_x'$, (A.2.12) follows from combining (A.2.3) and (A.2.11) and then applying a union bound. ■

***Lemma A.2.7*** *(bounding $\|h_0\|_2$) if conditions in lemma A.2.2 holds, $h_0$ is defined in (A.2.1), one has,*

$$prob\left(\|h_0\|_2 \le \sqrt{|s_f|} + \sqrt{\frac{m}{|s_f^c|}}\sqrt{6}\left(\lambda^{-1} + \sqrt{2ln(2|s_x|/\varepsilon)}\right)\sqrt{|s_x|}\right) \ge 1 - 3\varepsilon \quad \text{(A.2.13)}$$

Proof:

By the triangle inequality, one has,

$$\|h_0\|_2 \le \|h_0(s_f)\|_2 + \|h_0(s_f^c)\|_2 \quad \text{(A.2.14)}$$

The conclusion of the lemma follows from (A.2.1) and lemma A.2.6.■

Now we are in the position of bounding $\|q_0\|_2$, as stated in the below lemma A.2.8.

***Lemma A.2.8****(bounding $\|q_0\|_2$) If conditions in lemma A.2.2 hold, $h_0$ and $q_0$ are defined in (A.2.1) and (A.2.2), respectively, then,*

$$prob\left(\|q_0\|_2 \le \rho_1\sqrt{|s_f|} + \rho_2\sqrt{|s_x|}\right) \ge 1 - 3\varepsilon \quad \text{(A.2.15)}$$

*where* $\rho_1 = \sqrt{\frac{n}{m}}\lambda,\ \rho_2 = \sqrt{\frac{m}{|s_f^c|}}\sqrt{6}\left(\lambda^{-1} + \sqrt{2ln(2|s_x|/\varepsilon)}\right) + \sqrt{6}\left(1 + \lambda\sqrt{2ln(2|s_x|/\varepsilon)}\right)\sqrt{\frac{n}{|s_f^c|}}$.

Proof:

By triangle inequality,

$$\|q_0\|_2 \le \|q_0(1:|s_f^c|)\|_2 + \|q_0(|s_f^c| + 1: |s_f^c| + |s_x^c|)\|_2 \quad \text{(A.2.16)}$$

According to (A.2.2), $q_0(1:|s_f^c|) = h_0(s_f^c)$ and by lemma A.2.6, one has,

$$prob\left(\|q_0(1:|s_f^c|)\|_2 \le \sqrt{\frac{m}{|s_f^c|}}\sqrt{6}\left(\lambda^{-1} + \sqrt{2ln(2|s_x|/\varepsilon)}\right)\sqrt{|s_x|}\right) \ge 1 - 3\varepsilon \quad \text{(A.2.17)}$$

since $q_0(|s_f^c| + 1: |s_x^c| + |s_f^c|) = -\lambda\widetilde{A}^*(s_x^c, [m])h_0$ according to (A.2.2), using the fact that

$$\left\|-\lambda\widetilde{A}^*(s_x^c, [m])h_0\right\|_2 \le \lambda\sqrt{\frac{n}{m}}\|h_0\|_2 \quad \text{(A.2.18)}$$

By lemma A.2.7, one has,

$$prob\left(\|q_0(|s_f^c| + 1: |s_x^c| + |s_f^c|)\|_2 \le \sqrt{\frac{n}{m}}\left(\lambda\sqrt{|s_f|} + \sqrt{\frac{m}{|s_f^c|}}\sqrt{6}\left(1 + \lambda\sqrt{2ln(2|s_x|/\varepsilon)}\right)\sqrt{|s_x|}\right)\right) \ge 1 - 3\varepsilon \quad \text{(A.2.19)}$$

combining (A.2.17) and (A.2.19) prove the conclusion of the lemma.■

***Theorem A.2.1*** *if conditions in lemma A.2.2 hold, let $q_0$ be defined in (A.2.2) which satisfies $\Phi q_0 = w$, if*

$$\rho_1\sqrt{|s_f|} + \rho_2\sqrt{|s_x|} \le \frac{1}{2}\left(1 - \xi_k(\Phi)\right)\sqrt{k} \quad \text{(A.2.20)}$$

*Where $k = |s_f^c| - |s_x|$, $\rho_1$, $\rho_2$ are defined in lemma A.2.8. Then one has,*

$$prob\left(\|q_0\|_2 + \frac{\xi_k(\Phi)}{1-\xi_k(\Phi)}\|\mathcal{S}[q_0]\|_2 \le \sqrt{k}/2\right) \ge 1 - 3\varepsilon \quad \text{(A.2.21)}$$

*where $0 < \varepsilon < \frac{1}{3}$ is a constant.*

Proof:

According to theorem A.1.1, one has $0 < \xi_k(\Phi) < 1$, since $\|\mathcal{S}[q_0]\|_2 \le \|q_0\|_2$, to show $\|q_0\|_2 + \frac{\xi_k(\Phi)}{1-\xi_k(\Phi)}\|\mathcal{S}[q_0]\|_2 \le \sqrt{k}/2$, it is sufficient to show $\|q_0\|_2 \le \frac{1-\xi_k(\Phi)}{2}\sqrt{k}$. (A.2.21) follows immediately from (A.2.20) and (A.2.15) in lemma A.2.8.■

## A.3 proving that B in lemma 3.1 is full rank

***Theorem A.3.1*** *if* $\wedge$ *and* $\wedge(s_f)$ *are random subsets of* $[n]$, $|s_f^c| \geq \frac{32}{3}|s_x| ln\left(\frac{2|s_x|}{\varepsilon}\right)$, *then with probability at least* $1-\varepsilon$, *matrix* $B = [\lambda\tilde{A}([m], s_x), I_m([m], s_f)]$ *is full rank.*

Proof: to show B is full rank, it's sufficient to show that matrix $C = B^*B$ is non-singular, in other words, the minimum eigenvalue of C is positive. Since,

$$C = \begin{bmatrix} \lambda^2\tilde{A}^*(s_x, [m])\tilde{A}([m], s_x) & \lambda\tilde{A}^*(s_x, s_f) \\ \lambda\tilde{A}(s_f, s_x) & I_{|s_f|} \end{bmatrix} \quad \text{(A.3.1)}$$

By Schur complement decomposition, one has,

$$HCH^T = \begin{bmatrix} \lambda^2\tilde{A}^*(s_x, s_f^c)\tilde{A}(s_f^c, s_x) & 0 \\ 0 & I_{|s_f|} \end{bmatrix} \quad \text{(A.3.2)}$$

Where $H = \begin{bmatrix} I_{|s_x|} & -\lambda\tilde{A}^*(s_x, s_f) \\ 0 & I_{|s_f|} \end{bmatrix}$ is a non-singular matrix, by lemma A.2.1, one has

$$\text{prob}\left(\left\|\lambda^2\tilde{A}^*(s_x, s_f^c)\tilde{A}(s_f^c, s_x)\right\|_{2\to2} \geq \frac{\lambda^2}{2}\right) \geq 1-\varepsilon \quad \text{(A.3.3)}$$

Holds if $|s_f^c| \geq \frac{32}{3}|s_x| ln\left(\frac{2|s_x|}{\varepsilon}\right)$.

The conclusion of this theorem is proved by combining (A.3.2) and (A.3.3). ■

## A.4 Proof of the theorem 2.1

Proof of theorem 2.1:

Firstly, matrix $B = [\lambda\tilde{A}([m], s_x), I_m([m], s_f)]$ is proved to be full rank with probability at least $1-\varepsilon$ in theorem A.3.1.

Furthermore, defining $q_0$ as in (A.2.2), let $k = |s_f^c| - |s_x|$ and $c = 1 - \xi_k(\Phi)$ then according to theorem A.2.1, we conclude that $q_0$ satisfies (3.8) with probability at least $1-3\varepsilon$, which proves the conclusion of the theorem according to lemma 3.1~lemma 3.3.■